\documentclass[a4paper,12pt]{article}%
\usepackage{amsmath}
\usepackage{amsfonts}
\usepackage{amssymb}
\usepackage{graphicx}%
\usepackage{epsf,epsfig,graphics,epstopdf}
\begin{document}

\title{Introduction to Renomalization in Field Theory}
\author{Ling-Fong Li\\Carnegie Mellon University, Pittsburgh, PA. USA\\and \\Chongqing Univerisity of Posts \& Telecommunications, \\Chongqing, China}
\date{}
\maketitle
\tableofcontents

\begin{abstract}
A simple introduction of renormalization in quantum field theory is discussed.
Explanation of concepts is emphasized instead of the technical details.

\end{abstract}

\section{Introduction}

Many people who have studied quantum field theory find the most difficult part
is the theory of renormalization. The relativistic field theory is full of
infinities which need to be taken care of before the theoretical predictions
can be compared with experimental measurements. These infinities look
formidable at first sight. It is remarkable that over the years a way has been
found to make sense of these apparently divergent theories
(\cite{Renormalization},\cite{Renormalization-advanced}).

The theory of renormalization is a prescription which consistently isolates
and removes all these infinities from the physically measurable quantities.
Note that the need for renomalization is quite general and is not unique to
the relativistic field theory. For example, consider an electron moving inside
a solid. If the interaction between electron and the lattice of the solid is
weak enough, we can use an effective mass $m^{\ast}$ to describe its response
to an externally applied force and this effective mass is certainly different
from the mass $m$ measured outside the solid. Thus the electron mass is
changed (renormalized) from $m$ to $m^{\ast}$ by the interaction of the
electron with the lattice in the solid. In this simple case, both $m$ and
$m^{\ast}$ are measurable and hence finite. For the relativistic field theory,
the situation is the same except for two important differences. First the
renormalization due to the interaction is generally infinite (corresponding to
the divergent loop diagrams). These infinities, coming from the contribution
of high momentum modes are present even for the cases where the interactions
are weak. Second, there is no way to switch off the interaction between
particles and the quantities in the absence of interaction, bare quantities,
are not measurable. Roughly speaking, the program of removing the infinities
from physically measurable quantities in relativistic field theory, the
renormalization program, involves shuffling all the divergences into bare
quantities. In other words, we can redefine the unmeasurable quantities to
absorb the divergences so that the physically measurable quantities are
finite. The renormalized mass which is now finite can only be determined from
experimental measurement and can not be predicted from the theory alone.

Eventhough the concept of renormalization is quite simple, the actual
procedure for carrying out the operation is quite complicated and
intimidating. In this article, we will give a bare bone of this program and
refer interested readers to more advanced literature (\cite{BPH}%
,\cite{Zimmermann}). Note that we need to use some regularization procedure
(\cite{Dim reg}) to make these divergent quantities finite before we can do
mathematically meaningful manipulations. We will not discuss this part in the
short presentation here. Also note that not every relativistic field theory
will have this property that all divergences can be absorbed into redefinition
of few physical parameters. Those which have this property are called
$renormalizable$ theories and those which don't are called unrenormalizable
theories. This has became an important criteria for choosing a right theory
because we do not really know how to handle the $unrenormalizable$ theory.

\section{Renormalization Schemes}

There are two different methods to carry out the renormalization program, i)
conventional renormalization which is more intuitive but mathematically
complicated, ii) BPH renormalization which is simple to describe but not so
transparent (\cite{BPH}). These two methods are in fact complementary to each
other and it is very useful to know both.

\subsection{Conventional renormalization}

We will illustrate this scheme in the simple $\lambda\varphi^{4}$ theory where
the Lagrangian can be written as%
\[%
%TCIMACRO{\tciLaplace}%
%BeginExpansion
\mathcal{L}%
%EndExpansion
=%
%TCIMACRO{\tciLaplace}%
%BeginExpansion
\mathcal{L}%
%EndExpansion
_{0}+%
%TCIMACRO{\tciLaplace}%
%BeginExpansion
\mathcal{L}%
%EndExpansion
_{1}%
\]
with%
\[%
%TCIMACRO{\tciLaplace}%
%BeginExpansion
\mathcal{L}%
%EndExpansion
_{0}=\dfrac{1}{2}\left[  \left(  \partial_{\mu}\varphi_{0}\right)  ^{2}%
-\mu_{0}^{2}\varphi_{0}^{2}\right]
\]
and%
\[%
%TCIMACRO{\tciLaplace}%
%BeginExpansion
\mathcal{L}%
%EndExpansion
_{1}=-\dfrac{\lambda_{0}}{4!}\varphi_{0}^{4}%
\]
Here $\mu_{0}$\ ,$\lambda_{0},\varphi_{0}$ are bare mass, bare coupling
constant and bare field respectively. The propagator and vertex of this theory
are given below,
\begin{figure}
\begin{center}
{\includegraphics{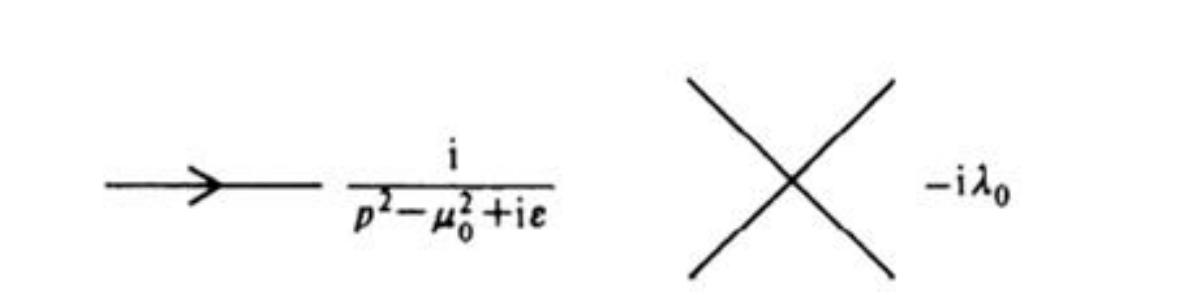}}
\\
Fig 1 Feynman rule for $\lambda\phi^{4}$ theory
\end{center}
\end{figure}
Here $p$ is the momentum carried by the line and $\mu_{0}^{2}$ is the bare
mass term in $%
%TCIMACRO{\tciLaplace}%
%BeginExpansion
\mathcal{L}%
%EndExpansion
_{0}.$

The two point function (propagator) defined by%
\[
i\Delta\left(  p\right)  =\int d^{4}xe^{-ip\cdot x}\left\langle 0\left\vert
T\left(  \varphi_{0}\left(  x\right)  \varphi_{0}\left(  0\right)  \right)
\right\vert 0\right\rangle
\]
can be written in terms of one-particle-irreducible, or 1PI ( those graphs
which can not be made disconnected by cutting any one internal line) as a
geometric series%
\begin{align}
i\Delta\left(  p\right)   &  =\dfrac{i}{p^{2}-\mu_{0}^{2}+i\varepsilon}%
+\dfrac{i}{p^{2}-\mu_{0}^{2}+i\varepsilon}\left(  -i\Sigma\left(
p^{2}\right)  \right)  \dfrac{i}{p^{2}-\mu_{0}^{2}+i\varepsilon}%
+\cdots\label{Propagator}\\
&  =\dfrac{i}{p^{2}-\mu_{0}^{2}-\Sigma\left(  p^{2}\right)  +i\varepsilon
}\nonumber
\end{align}
Here $\Sigma\left(  p^{2}\right)  $ is the IPI self energy graph. In one-loop,
the 1PI divergent graphs are
\begin{figure}
\begin{center}
{\includegraphics{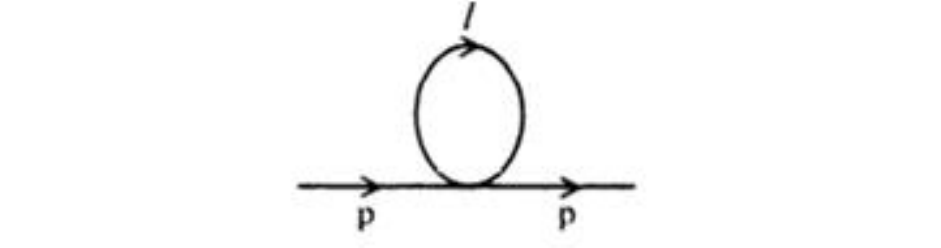}}
\\
Fig 2. 1-loop 2-point funcrion
\end{center}
\end{figure}

\begin{figure}
\begin{center}
{\includegraphics{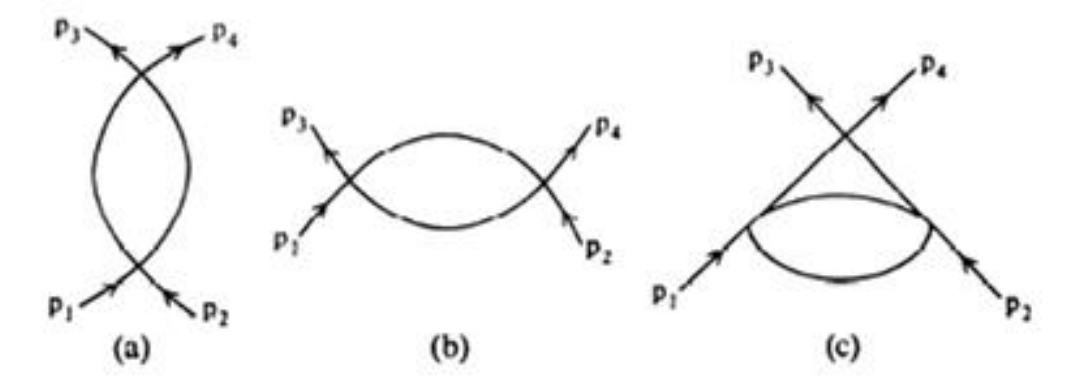}}
\\
Fig.3 One-loop 4-point functions
\end{center}
\end{figure}
For the self energy the contribution is,%
\begin{equation}
-i\Sigma\left(  p^{2}\right)  =-\dfrac{i\lambda_{0}}{2}\int\dfrac{d^{4}%
l}{\left(  2\pi\right)  ^{4}}\dfrac{i}{l^{2}-\mu_{0}^{2}+i\varepsilon
}\label{self-energy}%
\end{equation}
which diverges quadratically and for the 4-point functions we have%
\begin{equation}
\Gamma_{a}=\Gamma\left(  p^{2}\right)  =\Gamma\left(  s\right)  =\dfrac
{\left(  -i\lambda_{0}\right)  ^{2}}{2}\int\dfrac{d^{4}l}{\left(  2\pi\right)
^{4}}\dfrac{i}{l^{2}-\mu_{0}^{2}+i\varepsilon}\dfrac{i}{\left(  l-p\right)
^{2}-\mu_{0}^{2}+i\varepsilon}\label{4-point fn}%
\end{equation}%
\[
\Gamma_{b}=\Gamma\left(  t\right)  ,\qquad\Gamma_{c}=\Gamma\left(  u\right)
\]
Here
\[
s=p^{2}=\left(  p_{1}+p_{2}\right)  ^{2},\qquad t=\left(  p_{1}-p_{3}\right)
^{2},\qquad u=\left(  p_{1}-p_{4}\right)  ^{2},
\]
are the Mandelstam variables and $\Gamma\left(  s\right)  $ diverges logarithmically.

One important feature to note about these integrals is that when we
differentiate them with respect to external momenta, the integral will become
more convergent. For example, if we differentiate $\Gamma\left(  p^{2}\right)
$ with respect to $p^{2}$ , one finds%
\begin{align*}
\dfrac{\partial}{\partial p^{2}}\Gamma\left(  p^{2}\right)   &  =\dfrac
{1}{2p^{2}}p_{\mu}\dfrac{\partial}{\partial p_{\mu}}\Gamma\left(  p^{2}\right)
\\
&  =\dfrac{\lambda_{0}^{2}}{p^{2}}\int\dfrac{d^{4}l}{\left(  2\pi\right)
^{4}}\dfrac{\left(  l-p\right)  \cdot p}{l^{2}-\mu_{0}^{2}+i\varepsilon}%
\dfrac{1}{\left[  \left(  l-p\right)  ^{2}-\mu_{0}^{2}+i\varepsilon\right]
^{2}}%
\end{align*}
which is finite. This means that the divergences will reside only in the first
few terms in a Taylor expansion in the external momenta of the Feynman
diagram. In our case, we can write%
\[
\Gamma\left(  s\right)  =\Gamma\left(  0\right)  +\overset{\_}{\Gamma}\left(
s\right)
\]
where $\Gamma\left(  0\right)  $ is logarithmic divergent and $\overset
{\_}{\Gamma}\left(  s\right)  ,$ which is the sum of all higher derivative
terms, is finite. In other words, the finite part $\overset{\_}{\Gamma}\left(
s\right)  $ corresponds to subtracting the divergent part $\Gamma\left(
0\right)  $ from $\Gamma\left(  s\right)  $ and is sometimes referred to as
the $substraction$.\newline\underline{\textbf{Mass and wavefunction
renormalization}}

The self energy contribution in Eq (\ref{self-energy}) is quadratically
divergent. To isolate the divergences we use the Taylor expansion around some
arbitrary value $\mu^{2},$%
\[
\Sigma\left(  p^{2}\right)  =\Sigma\left(  \mu^{2}\right)  +\left(  p^{2}%
-\mu^{2}\right)  \Sigma^{\prime}\left(  \mu^{2}\right)  +\widetilde{\Sigma
}\left(  p^{2}\right)
\]
where $\Sigma\left(  \mu^{2}\right)  $ is quadratically divergent, and
$\Sigma^{\prime}\left(  \mu^{2}\right)  $ is logarithmically divergent and
$\widetilde{\Sigma}\left(  p^{2}\right)  $ is finite. The finite part
$\widetilde{\Sigma}\left(  p^{2}\right)  $ will have the property,%
\begin{equation}
\widetilde{\Sigma}\left(  \mu^{2}\right)  =\widetilde{\Sigma}^{\prime}\left(
\mu^{2}\right)  =0 \label{finite-part}%
\end{equation}
Note that self-energy in 1-loop has the peculiar feature that it is
independent of the external momentum $p^{2}$ and the Taylor expansion has only
one term, $\Sigma\left(  \mu^{2}\right)  .$ However, the higher loop
contribution does depend on the external momentum and the Taylor expansion is
non-trivial. The propagator in Eq (\ref{Propagator}) is then,%
\[
i\Delta\left(  p\right)  =\dfrac{i}{p^{2}-\mu_{0}^{2}-\Sigma\left(  \mu
^{2}\right)  -\left(  p^{2}-\mu^{2}\right)  \Sigma^{\prime}\left(  \mu
^{2}\right)  -\widetilde{\Sigma}\left(  p^{2}\right)  +i\varepsilon}%
\]
The physical mass is defined as the position of the pole in the propagator.
Since up to this point $\mu^{2}$ is arbitrary, we can choose it to satisfy the
relation,%
\begin{equation}
\mu_{0}^{2}+\Sigma\left(  \mu^{2}\right)  =\mu^{2},\qquad
\label{mass-renormalization}%
\end{equation}
Then
\[
i\Delta\left(  p\right)  =\dfrac{i}{\left(  p^{2}-\mu^{2}\right)  \left[
1-\Sigma^{\prime}\left(  \mu^{2}\right)  \right]  -\widetilde{\Sigma}\left(
p^{2}\right)  +i\varepsilon}%
\]
and using Eq (\ref{finite-part}) we see that $\Delta\left(  p\right)  $ has a
pole at $p^{2}=\mu^{2}.$ Thus $\mu^{2}$ is the \textbf{physical mass} and is
related to the bare mass $\mu_{0}^{2}$ in Eq (\ref{mass-renormalization}).
This is the $mass$ $renormalization.$ Since $\Sigma\left(  \mu^{2}\right)  $
is divergent, the bare mass $\mu_{0}^{2}$ must also be divergent so that the
combination $\mu_{0}^{2}+\Sigma\left(  \mu^{2}\right)  $ is finite and
measurable. In other words, the bare mass $\mu_{0}^{2}$ has to diverge in such
a way that its divergence cancels the divergent loop correction to yield a
finite result. It amounts to shuffling the infinities to unobservable
quantities like bare mass $\mu_{0}^{2}.$ This is the part in renormalization
theory which is very difficult to comprehend at the first sight. Nevertheless
it is logically consistent and the rules are very precise. Furthermore, the
results after the renomalization have been successfully checked by
experiments. This gives us confidence about the validity of renormalization theory.

To remove the divergent quantity $\Sigma^{\prime}\left(  \mu^{2}\right)  $ we
note that in 1-loop both $\Sigma^{\prime}\left(  \mu^{2}\right)
,\widetilde{\Sigma}\left(  p^{2}\right)  $ are of order $\lambda_{0},$ for
convenience, we can make the approximation,%
\[
\widetilde{\Sigma}\left(  p^{2}\right)  \simeq\left[  1-\Sigma^{\prime}\left(
\mu^{2}\right)  \right]  \widetilde{\Sigma}\left(  p^{2}\right)  +O\left(
\lambda_{0}^{2}\right)
\]
and write the propagator as%
\[
i\Delta\left(  p\right)  =\dfrac{iZ_{\varphi}}{\left(  p^{2}-\mu^{2}\right)
-\widetilde{\Sigma}\left(  p^{2}\right)  +i\varepsilon}%
\]
where%
\begin{equation}
Z_{\varphi}=\left[  1-\Sigma^{\prime}\left(  \mu^{2}\right)  \right]
^{-1}\simeq1+\Sigma^{\prime}\left(  \mu^{2}\right)  +O\left(  \lambda_{0}%
^{2}\right)  \label{Zphi}%
\end{equation}
Now the divergence is shuffled into the multiplicative factor $Z_{\varphi}$
which can be removed by defining a renormalized field $\varphi$ as,%
\begin{equation}
\varphi=Z_{\varphi}^{-1/2}\varphi_{0} \label{Zfield}%
\end{equation}
The propagator for the renormalized field is then%
\begin{align}
i\Delta_{R}\left(  p\right)   &  =\int d^{4}xe^{-ip\cdot x}\left\langle
0\left\vert T\left(  \varphi\left(  x\right)  \varphi\left(  0\right)
\right)  \right\vert 0\right\rangle \label{remoralized propagator}\\
&  =iZ_{\varphi}^{-1}\Delta\left(  p\right)  =\dfrac{i}{\left(  p^{2}-\mu
^{2}\right)  -\widetilde{\Sigma}\left(  p^{2}\right)  +i\varepsilon}\nonumber
\end{align}
and it is completely finite. $Z_{\varphi}$ is usually called the
$wavefunction$ $renormalization$ $constant$. Thus another divergence is
shuffled into the bare field operator $\varphi_{0}$ which is also not measurable.

The new renormalized field operator $\varphi$ should also be applied to the
renormalized higher point Green's functions,%
\begin{align*}
G_{R}^{\left(  n\right)  }\left(  x_{1},x_{2},\cdots x_{n}\right)   &
=\left\langle 0\left\vert T\left(  \varphi\left(  x_{1}\right)  \varphi\left(
x_{2}\right)  \cdots\varphi\left(  x_{n}\right)  \right)  \right\vert
0\right\rangle \\
&  =Z_{\varphi}^{-n/2}\left\langle 0\left\vert T\left(  \varphi_{0}\left(
x_{1}\right)  \varphi_{0}\left(  x_{2}\right)  \cdots\varphi_{0}\left(
x_{n}\right)  \right)  \right\vert 0\right\rangle \\
&  =Z_{\varphi}^{-n/2}G_{0}^{\left(  n\right)  }\left(  x_{1},x_{2},\cdots
x_{n}\right)
\end{align*}
Here $G_{0}^{\left(  n\right)  }\left(  x_{1},x_{2},\cdots x_{n}\right)  $ is
the unrenormalized $n-$point Green's function. Or in momentum space%
\[
G_{R}^{\left(  n\right)  }\left(  p_{1},p_{2},\cdots p_{n}\right)
=Z_{\varphi}^{-n/2}G_{0}^{\left(  n\right)  }\left(  p_{1},p_{2},\cdots
p_{n}\right)
\]
where%
\[
\left(  2\pi\right)  ^{4}\delta^{4}\left(  p_{1}+\cdots p_{n}\right)
G_{R}^{\left(  n\right)  }\left(  p_{1},\cdots p_{n}\right)  =\int\left(
%TCIMACRO{\dprod \limits_{i=1}^{n}}%
%BeginExpansion
{\displaystyle\prod\limits_{i=1}^{n}}
%EndExpansion
dx_{i}^{4}e^{-ip_{i}\cdot x_{i}}\right)  G_{R}^{\left(  n\right)  }\left(
x_{1},\cdots x_{n}\right)
\]
Similarly for $G_{0}^{\left(  n\right)  }\left(  p_{1},p_{2},\cdots
p_{n}\right)  .$ To go from the connected Green's functions to the 1PI
(amputated) Green's functions, we need to eliminate the one-particle reducible
diagrams, and also to remove the propagators $i\Delta_{R}\left(  p_{i}\right)
$ for the external lines in 1PI Green's function $G_{R}^{\left(  n\right)
}\left(  p_{1},\cdots p_{n}\right)  .$ As a result the relation between 1PI
Green's functions are of the form,%
\[
\Gamma_{R}^{\left(  n\right)  }\left(  p_{1},p_{2},\cdots p_{n}\right)
=Z_{\varphi}^{n/2}\Gamma_{0}^{\left(  n\right)  }\left(  p_{1},p_{2},\cdots
p_{n}\right)
\]

Note that the relations in Eq (\ref{finite-part}) are direct consequence of
the Taylor expansion around the point $p^{2}=\mu^{2}$ which is totally
arbitrary. From the form of the renormalized propagator in Eq
(\ref{remoralized propagator}), we see that Eq (\ref{finite-part}) are
equivalent to the relations%
\[
\Delta_{R}^{-1}\left(  \mu^{2}\right)  =0,\qquad\left.  \dfrac{d}{dp^{2}%
}\Delta_{R}^{-1}\left(  p^{2}\right)  \right\vert _{p^{2}=\mu^{2}}=1
\]
If we have chosen some other point, e.g. $p^{2}=0$ for the Taylor expansion,
the finite part $\widetilde{\Sigma}_{1}\left(  p^{2}\right)  $ will have the
properties
\begin{equation}
\widetilde{\Sigma}_{1}\left(  0\right)  =\widetilde{\Sigma}_{1}^{\prime
}\left(  0\right)  =0 \label{finite part 2}%
\end{equation}
Or in terms of renormalized propagator,%
\[
\Delta_{R}^{-1}\left(  0\right)  =-\mu^{2},\qquad\left.  \dfrac{d}{dp^{2}%
}\Delta_{R}^{-1}\left(  p^{2}\right)  \right\vert _{p^{2}=0}=1
\]
Sometimes in the renormalization prescription we replace the statement "
Taylor expansion around $p^{2}=\mu^{2},$or $p^{2}=0"$ by relations expressed,
in Eq (\ref{finite-part},\ref{finite part 2}), called the $renormalization$
$conditions$. One important feature to keep in mind is that in carrying out
the renormalization program there is an arbitrariness in choosing the points
for the Taylor expansion. Different renormalization schemes seem to give rise
to different looking relations. However, if these renormalization schemes make
any sense at all, the physical laws which are relations among physically
measurable quantities should be the same regardless of which scheme is used.
This is the basic idea behind the $renormalization$ $group$ $equations$
(\cite{Renor Group}).\newline\underline{\textbf{Coupling constant
renormalization}}

The basic coupling in $\lambda\varphi^{4}$ theory is the 4-point function
which in 1-loop has the form, before renormalization,%
\[
\Gamma_{0}^{\left(  4\right)  }\left(  s,t,u\right)  =-i\lambda_{0}%
+\Gamma\left(  s\right)  +\Gamma\left(  t\right)  +\Gamma\left(  u\right)
\]
where last three terms are logarithmic divergent. We will remove these
divergences by the redefinition of the coupling constant. Note that the
physical coupling constant is measured in terms of two-particle scattering
amplitude which is essentially 1PI 4-point Green's function $\Gamma
_{R}^{\left(  4\right)  }\left(  s,t,u\right)  $ which is a function of the
kinematical variables, $s,t$ and $u.$ For convenience, we can choose the
symmetric point,%
\[
s_{0}=t_{0}=u_{0}=\dfrac{4\mu^{2}}{3}%
\]
to define the coupling constant,%
\[
\Gamma_{R}^{\left(  4\right)  }\left(  s_{0},t_{0},u_{0}\right)  =-i\lambda
\]
where $\lambda$ is the renormalized coupling constant. Since $\Gamma\left(
s\right)  $ is only logarithmically divergent, we can isolate the divergence
in one term in the Taylor expansion,%
\[
\Gamma\left(  s\right)  =\Gamma\left(  s_{0}\right)  +\widetilde{\Gamma
}\left(  s\right)
\]
where $\widetilde{\Gamma}\left(  s\right)  $ is finite and
\[
\widetilde{\Gamma}\left(  s_{0}\right)  =0
\]
Then%
\[
\Gamma_{0}^{\left(  4\right)  }\left(  s,t,u\right)  =-i\lambda_{0}%
+3\Gamma\left(  s_{0}\right)  +\widetilde{\Gamma}\left(  s\right)
+\widetilde{\Gamma}\left(  t\right)  +\widetilde{\Gamma}\left(  u\right)
\]
We can isolate the divergence by combining the first two term and define the
vertex renormalization constant $Z_{\lambda}$ by%
\[
-iZ_{\lambda}\lambda_{0}=-i\lambda_{0}+3\Gamma\left(  s_{0}\right)
\]
Then%
\[
\Gamma_{0}^{\left(  4\right)  }\left(  s,t,u\right)  =-iZ_{\lambda}\lambda
_{0}+\widetilde{\Gamma}\left(  s\right)  +\widetilde{\Gamma}\left(  t\right)
+\widetilde{\Gamma}\left(  u\right)
\]
The renomalized 4-point 1PI is then%
\begin{align}
\Gamma_{R}^{\left(  4\right)  }\left(  s,t,u\right)   &  =Z_{\varphi}%
^{2}\Gamma_{0}^{\left(  4\right)  }\left(  s,t,u\right) \label{renom 1PI}\\
&  =-iZ_{\lambda}Z_{\varphi}^{2}\lambda_{0}+Z_{\varphi}^{2}\left[
\widetilde{\Gamma}\left(  s\right)  +\widetilde{\Gamma}\left(  t\right)
+\widetilde{\Gamma}\left(  u\right)  \right] \nonumber
\end{align}
We now define the renomalized coupling constant $\lambda$ as%
\begin{equation}
\lambda=Z_{\lambda}Z_{\varphi}^{2}\lambda_{0} \label{renormalized coupling}%
\end{equation}
and from Eq (\ref{Zphi}) we see that
\[
Z_{\varphi}=1+O\left(  \lambda_{0}\right)
\]
Also $\widetilde{\Gamma}$ is of order of $\lambda_{0}^{2}.$ The renomalized
4-point 1PI can be put into the form,%
\[
\Gamma_{R}^{\left(  4\right)  }\left(  s,t,u\right)  =\lambda+\left[
\widetilde{\Gamma}\left(  s\right)  +\widetilde{\Gamma}\left(  t\right)
+\widetilde{\Gamma}\left(  u\right)  \right]  +O\left(  \lambda_{0}%
^{3}\right)
\]
Assuming that the coupling constant $\lambda$ is measured in the scattering
experiment and is finite, we see that this 4-point function is completely free
of divergences. Eq (\ref{renormalized coupling}) shows that the
renormalization of coupling constant involves wavefunction renormalization in
addition to the vertex correction.

For the renormalization of connected Green's functions, we need to add
one-particle reducible diagrams and attach propagators for the external lines.
We want to show that the renormalized Green functions when expressed in terms
of renormalized quantities are completely finite. We start with the
unrenormalized Green's function of the form,%
\begin{align}
G_{0}^{\left(  4\right)  }\left(  p_{1},\cdots p_{4}\right)   &  =%
%TCIMACRO{\tprod \limits_{j=1}^{4}}%
%BeginExpansion
{\textstyle\prod\limits_{j=1}^{4}}
%EndExpansion
\Delta^{\left(  0\right)  }\left(  p_{j}\right)  \{-i\lambda_{0}%
+3\Gamma\left(  s_{0}\right)  +\widetilde{\Gamma}\left(  s\right)
+\widetilde{\Gamma}\left(  t\right)  +\widetilde{\Gamma}\left(  u\right)
\label{unrenor Green}\\
&  +\left(  -i\lambda_{0}\right)  \sum_{k=1}^{4}\left[  -i\Sigma\left(
p_{k}^{2}\right)  i\Delta^{\left(  0\right)  }\left(  p_{k}\right)  \right]
\}
\end{align}
where
\[
\Delta^{\left(  0\right)  }\left(  p_{j}\right)  =\dfrac{1}{p_{j}^{2}-\mu
_{0}^{2}+i\varepsilon}%
\]
is the zeroth order bare propagator and the last terms here are coming from
the diagrams of the following type,
\begin{figure}
\begin{center}
{\includegraphics{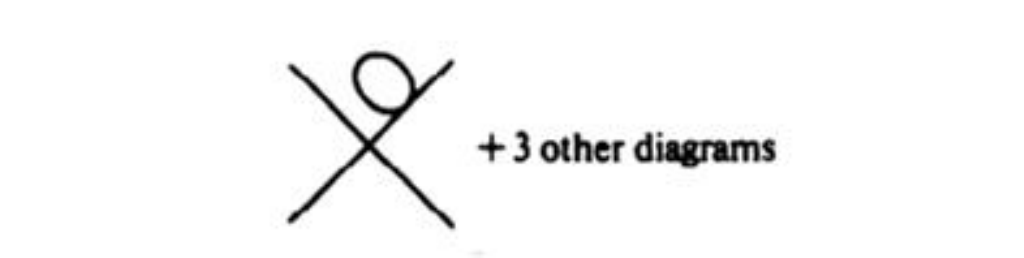}}
\\
Fig 4, 1-particle reducible 4 point function
\end{center}
\end{figure}

We can combine the first term and the last terms in $G_{0}^{\left(  4\right)
}\left(  p_{1},\cdots p_{4}\right)  $ to get%
\begin{align*}
&  \left(  -i\lambda_{0}\right)  \left\{  1+\sum_{k=1}^{4}\left[
\Sigma\left(  p_{k}^{2}\right)  \Delta^{\left(  0\right)  }\left(
p_{k}\right)  \right]  \right\}  \simeq\left(  -i\lambda_{0}\right)  \left[
\prod\limits_{k=1}^{4}\dfrac{1}{1-\Sigma\left(  p_{k}^{2}\right)
\Delta^{\left(  0\right)  }\left(  p_{k}\right)  }\right]  +O\left(
\lambda_{0}^{3}\right) \\
&  =\left(  -i\lambda_{0}\right)  \prod\limits_{k=1}^{4}\{\left[
\Delta^{\left(  0\right)  }\left(  p_{k}\right)  \right]  ^{-1}\dfrac
{1}{\left[  p_{k}^{2}-\mu_{0}^{2}-\Sigma\left(  p_{k}^{2}\right)  \right]
}\}=\left(  -i\lambda_{0}\right)  \prod\limits_{k=1}^{4}\{\left[
\Delta^{\left(  0\right)  }\left(  p_{k}\right)  \right]  ^{-1}\Delta\left(
p_{k}\right)
\end{align*}
where%
\[
\Delta\left(  p_{k}\right)  =\dfrac{1}{\left[  p_{k}^{2}-\mu_{0}^{2}%
-\Sigma\left(  p_{k}^{2}\right)  \right]  }%
\]
Since the difference between $\Delta\left(  p_{k}\right)  $ and $\Delta
^{\left(  0\right)  }\left(  p_{k}\right)  $ is higher order in $\lambda_{0},$
we can make the approximation for the rest of the terms in Eq
(\ref{unrenor Green}),
\[%
%TCIMACRO{\tprod \limits_{j=1}^{4}}%
%BeginExpansion
{\textstyle\prod\limits_{j=1}^{4}}
%EndExpansion
\Delta^{\left(  0\right)  }\left(  p_{j}\right)  \left[  3\Gamma\left(
s_{0}\right)  +\widetilde{\Gamma}\left(  s\right)  +\widetilde{\Gamma}\left(
t\right)  +\widetilde{\Gamma}\left(  u\right)  \right]  \simeq%
%TCIMACRO{\tprod \limits_{j=1}^{4}}%
%BeginExpansion
{\textstyle\prod\limits_{j=1}^{4}}
%EndExpansion
\Delta\left(  p_{j}\right)  \left[  3\Gamma\left(  s_{0}\right)
+\widetilde{\Gamma}\left(  s\right)  +\widetilde{\Gamma}\left(  t\right)
+\widetilde{\Gamma}\left(  u\right)  \right]
\]
The unrenormalized Green's function is then%
\begin{align*}
G_{0}^{\left(  4\right)  }\left(  p_{1},\cdots p_{4}\right)   &  =\left[
%TCIMACRO{\tprod \limits_{j=1}^{4}}%
%BeginExpansion
{\textstyle\prod\limits_{j=1}^{4}}
%EndExpansion
\Delta\left(  p_{j}\right)  \right]  \left[  -i\lambda_{0}+3\Gamma\left(
s_{0}\right)  +\widetilde{\Gamma}\left(  s\right)  +\widetilde{\Gamma}\left(
t\right)  +\widetilde{\Gamma}\left(  u\right)  \right] \\
&  =\left[
%TCIMACRO{\tprod \limits_{j=1}^{4}}%
%BeginExpansion
{\textstyle\prod\limits_{j=1}^{4}}
%EndExpansion
\Delta\left(  p_{j}\right)  \right]  \Gamma_{0}^{\left(  4\right)  }\left(
s,t,u\right)
\end{align*}
We now multiply the unrenormalized Green's function by the appropriate factor
of $Z_{\varphi}$ to get the renomalized one,%
\begin{align*}
G_{R}^{\left(  4\right)  }\left(  p_{1},\cdots p_{4}\right)   &  =Z_{\varphi
}^{-2}G_{0}^{\left(  4\right)  }\left(  p_{1},\cdots p_{4}\right)
=Z_{\varphi}^{-2}\left[
%TCIMACRO{\tprod \limits_{j=1}^{4}}%
%BeginExpansion
{\textstyle\prod\limits_{j=1}^{4}}
%EndExpansion
\Delta\left(  p_{j}\right)  \right]  \Gamma_{0}^{\left(  4\right)  }\left(
s,t,u\right) \\
&  =Z_{\varphi}^{-2}\left[  Z_{\varphi}^{4}%
%TCIMACRO{\tprod \limits_{j=1}^{4}}%
%BeginExpansion
{\textstyle\prod\limits_{j=1}^{4}}
%EndExpansion
i\Delta_{R}\left(  p_{j}\right)  \right]  Z_{\varphi}^{-2}\Gamma_{R}^{\left(
4\right)  }\left(  s,t,u\right) \\
&  =\left[
%TCIMACRO{\tprod \limits_{j=1}^{4}}%
%BeginExpansion
{\textstyle\prod\limits_{j=1}^{4}}
%EndExpansion
i\Delta_{R}\left(  p_{j}\right)  \right]  \Gamma_{R}^{\left(  4\right)
}\left(  s,t,u\right)
\end{align*}
Thus we have removed all the divergences in the connected 4-point Green's function.

In summary, Green's functions can be made finite if we express the bare
quantities in terms of the renormalized ones through the relations,%
\begin{equation}
\varphi=Z_{\varphi}^{-1/2}\varphi_{0},\qquad\lambda=Z_{\lambda}^{-1}%
Z_{\varphi}^{2}\lambda_{0},\qquad\mu^{2}=\mu_{0}^{2}+\delta\mu^{2}
\label{renormalized}%
\end{equation}
where $\delta\mu^{2}=\Sigma\left(  \mu^{2}\right)  .$ More specifically, for
an n-point Green's function when we express the bare mass $\mu_{0}$ and bare
coupling $\lambda_{0}$ in terms of the renormalized mass $\mu$ and coupling
\ $\lambda,$ and multiply by $Z_{\varphi}^{-1/2}$ for each external line the
result (the renormalized $n-$point Green's function) is completely finite,%
\[
G_{R}^{\left(  n\right)  }\left(  p_{1},\cdots p_{n};\lambda,\mu\right)
=Z_{\varphi}^{-n/2}G_{0}^{\left(  n\right)  }\left(  p_{1},\cdots
p_{n};\lambda_{0},\mu_{0},\Lambda\right)
\]
where $\Lambda$ is the cutoff needed to define the divergent integrals. This
feature, in which all the divergences, after rewriting $\mu_{0}$ and
$\lambda_{0}$ in terms of $\mu$ and $\lambda$ are aggregated into some
multiplicative constants, is called being $multiplicatively$ $renormalizable$.

\bigskip

Our discussion here contains some of the essential features in the
renormalization program. To prove that the procedure we outline here will
remove all the divergences in the theory is a very complicated mathematical
undertaking and is beyond the scope of this simple introduction.

\subsection{BPH renormalization}

BPH renormalization ( Bogoliubov and Parasiuk, Hepp) (\cite{BPH}) is
completely equivalent to the conventional renormalization but organized
differently. We will illustrate this in the simple $\lambda\varphi^{4}$ theory.

Start from the unrenormalized Lagrangian,%
\[%
%TCIMACRO{\tciLaplace}%
%BeginExpansion
\mathcal{L}%
%EndExpansion
_{0}=\dfrac{1}{2}\left[  \left(  \partial_{\mu}\varphi_{0}\right)  ^{2}%
-\mu_{0}^{2}\varphi_{0}^{2}\right]  -\dfrac{\lambda_{0}}{4!}\varphi_{0}^{4}%
\]
where all the quantities are unrenormalized. We can rewrite this in terms of
renormalized quantities using Eq (\ref{renormalized}),%
\[%
%TCIMACRO{\tciLaplace}%
%BeginExpansion
\mathcal{L}%
%EndExpansion
_{0}=%
%TCIMACRO{\tciLaplace}%
%BeginExpansion
\mathcal{L}%
%EndExpansion
+\Delta%
%TCIMACRO{\tciLaplace}%
%BeginExpansion
\mathcal{L}%
%EndExpansion
\]
where%
\begin{equation}%
%TCIMACRO{\tciLaplace}%
%BeginExpansion
\mathcal{L}%
%EndExpansion
==\dfrac{1}{2}\left[  \left(  \partial_{\mu}\varphi\right)  ^{2}-\mu
^{2}\varphi^{2}\right]  -\dfrac{\lambda}{4!}\varphi^{4}
\label{renormalized Lagrangian}%
\end{equation}
has exactly the same form as the original Lagrangian, is called the
$renormalized$ $Lagrangian$, and%
\begin{equation}
\Delta%
%TCIMACRO{\tciLaplace}%
%BeginExpansion
\mathcal{L}%
%EndExpansion
=\dfrac{\left(  Z_{\varphi}-1\right)  }{2}\left[  \left(  \partial_{\mu
}\varphi\right)  ^{2}-\mu^{2}\varphi^{2}\right]  +\dfrac{\delta\mu^{2}}%
{2}Z_{\varphi}\varphi^{2}-\dfrac{\lambda\left(  Z_{\lambda}-1\right)  }%
{4!}\varphi^{4} \label{counterterm}%
\end{equation}
contains all the divergent constants, $Z_{\varphi},$ $Z_{\lambda},$ and
$\delta\mu^{2},$ and is called the counterterm Lagrangian.

The BPH renormalization scheme consists of the following steps;

\begin{enumerate}
\item Start with renormalized Lagrangian given in Eq
(\ref{renormalized Lagrangian}) to construct propagators and vertices.

\item Isolate the divergent parts of 1PI diagrams by Taylor expansion.
Construct a set of counterterms $\Delta%
%TCIMACRO{\tciLaplace}%
%BeginExpansion
\mathcal{L}%
%EndExpansion
^{\left(  1\right)  }$ which is designed to cancel these one-loop divergences.

\item A new Lagrangian $%
%TCIMACRO{\tciLaplace}%
%BeginExpansion
\mathcal{L}%
%EndExpansion
^{\left(  1\right)  }=%
%TCIMACRO{\tciLaplace}%
%BeginExpansion
\mathcal{L}%
%EndExpansion
+\Delta%
%TCIMACRO{\tciLaplace}%
%BeginExpansion
\mathcal{L}%
%EndExpansion
^{\left(  1\right)  }$ is used to generate the 2-loop diagrams and to
construct the counterterms $\Delta%
%TCIMACRO{\tciLaplace}%
%BeginExpansion
\mathcal{L}%
%EndExpansion
^{\left(  2\right)  }$ which cancels the divergences up to this order and so
on, as this sequence of operations is iteratively applied.
\end{enumerate}

The resulting Lagrangain is of the form,%
\[%
%TCIMACRO{\tciLaplace}%
%BeginExpansion
\mathcal{L}%
%EndExpansion
^{\left(  \infty\right)  }=%
%TCIMACRO{\tciLaplace}%
%BeginExpansion
\mathcal{L}%
%EndExpansion
+\Delta
\]

where the counterterm Lagrangian $\Delta%
%TCIMACRO{\tciLaplace}%
%BeginExpansion
\mathcal{L}%
%EndExpansion
$ is given by,%
\[
\Delta%
%TCIMACRO{\tciLaplace}%
%BeginExpansion
\mathcal{L}%
%EndExpansion
=\Delta%
%TCIMACRO{\tciLaplace}%
%BeginExpansion
\mathcal{L}%
%EndExpansion
^{\left(  1\right)  }+\Delta%
%TCIMACRO{\tciLaplace}%
%BeginExpansion
\mathcal{L}%
%EndExpansion
^{\left(  2\right)  }+\cdots\Delta%
%TCIMACRO{\tciLaplace}%
%BeginExpansion
\mathcal{L}%
%EndExpansion
^{\left(  n\right)  }+\cdots
\]
We will now show that\ the counter term Lagrangian has the same structure as
that in Eq ( \ref{counterterm}).\newline\underline{\textbf{Power Counting
Method}}

This method will help to classify divergences systematically. For a given
Feynamn diagram, we define \textbf{superficial degree of divergence }$D$ as
the number of loop momenta in the numerator minus the number of loop momenta
in the denominator. For illustration we will compute $D$ in $\lambda\phi^{4}$
theory. Define%
\[%
\begin{array}
[c]{l}%
B=\text{number of external lines}\\
IB=\text{number of internal lines}\\
n=\text{number of vertices}%
\end{array}
\]
It is straightforward to see that the superficial degree of divergence is
given by%
\begin{equation}
D=4-B \label{Superficial Div}%
\end{equation}
It is important to note that $D$ depends only on the number of external lines,
$B$ and not on $n,$ the number of vertices. This is a consequence of
$\lambda\phi^{4}$ theory and might not hold for other interactions. From this
Eq (\ref{Superficial Div}) we see that $D\geq0$ only for $B=2,4$ ( $B=$ even
because of the symmetry $\phi\rightarrow-\phi$ ). In the analysis of
divergences, we will use the superficial degree of divergences to construct
the counterterms. The reason for this will be explained later.

\begin{enumerate}
\item $B=2,\Rightarrow$ $D=2$\newline Being quadratically divergent, the
necessary Taylor expansion for the 2-point function is of the form,%
\[
\Sigma\left(  p^{2}\right)  =\Sigma\left(  0\right)  +p^{2}\Sigma^{\prime
}\left(  0\right)  +\widetilde{\Sigma}\left(  p^{2}\right)
\]
where $\Sigma\left(  0\right)  $ and $\Sigma^{\prime}\left(  0\right)  $ are
divergent and $\widetilde{\Sigma}\left(  p^{2}\right)  .$ To cancel these
divergences we need to add two counterterms,%
\[
\dfrac{1}{2}\Sigma\left(  0\right)  \phi^{2}+\dfrac{1}{2}\Sigma^{\prime
}\left(  0\right)  \left(  \partial_{\mu}\phi\right)  ^{2}%
\]
which give the following contributions,
\begin{figure}
\begin{center}
{\includegraphics{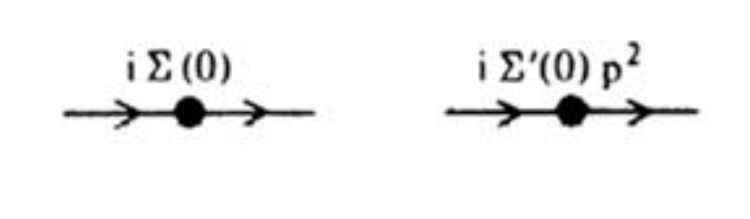}}
\\
Fig 5, counter terms for 2-point function
\end{center}
\end{figure}

\item $B=4,$ $\Rightarrow$ $\ D=0$\newline The Taylor expansion is%
\[
\Gamma^{\left(  4\right)  }\left(  p_{i}\right)  =\Gamma^{\left(  4\right)
}\left(  0\right)  +\widetilde{\Gamma}^{\left(  4\right)  }\left(
p_{i}\right)
\]
where $\Gamma^{\left(  4\right)  }\left(  0\right)  $ is logarithmically
divergent which is to be cancelled by conunterterm of the form%
\[
\dfrac{i}{4!}\Gamma^{\left(  4\right)  }\left(  0\right)  \phi^{4}%
\]
\begin{figure}
\begin{center}
{\includegraphics{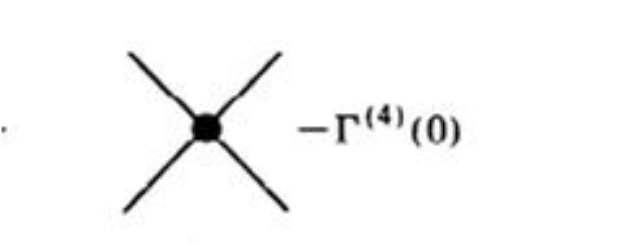}}
\\
Fig 6 counterterms for 4-point function
\end{center}
\end{figure}
\end{enumerate}

The general counterterrm Lagrangian is then%
\[
\Delta%
%TCIMACRO{\tciLaplace}%
%BeginExpansion
\mathcal{L}%
%EndExpansion
=\dfrac{1}{2}\Sigma\left(  0\right)  \phi^{2}+\dfrac{1}{2}\Sigma^{\prime
}\left(  0\right)  \left(  \partial_{\mu}\phi\right)  ^{2}+\dfrac{i}{4!}%
\Gamma^{\left(  4\right)  }\left(  0\right)  \phi^{4}%
\]
which is clearly the same as Eq(\ref{counterterm}) with the identification%
\[%
\begin{array}
[c]{l}%
\Sigma^{\prime}\left(  0\right)  =\left(  Z_{\varphi}-1\right) \\
\Sigma\left(  0\right)  =-\left(  Z_{\varphi}-1\right)  \mu^{2}+\delta\mu
^{2}\\
\Gamma^{\left(  4\right)  }\left(  0\right)  =-i\lambda\left(  1-Z_{\lambda
}\right)
\end{array}
\]
This illustrates the equivalence of BPH renormalization and conventional
renormalization.\newline\underline{\textbf{More on BPH\ renormalization}}

The BPH renormalization scheme looks very simple. It is remarkable that this
simple scheme can serve as the basis for setting up a proof for a certain
class of field theory. There are many interesting and useful features in BPH
which do not show themselves on the first glance and are very useful in the
understanding of this renormalization program. We will now discuss some of them.

\begin{enumerate}
\item \underline{\textbf{Convergence of Feynman diagrams}}

In our analysis so far, we have used the superficial degree of divergences $D
$. It is clear that to 1-loop order that superficial degree of divergence is
the same as the real degree of divergence. When we go beyond 1-loop it is
possible to have an overall $D<0$ while there are real divergences in the
subgraphs. The real convergence of a Feynman graph is governed by Weinberg's
theorem (\cite{Weinberg theorem}) : The general Feynman integral converges if
the superficial degree of divergence of the graph together with the
superficial degree of divergence of all subgraphs are negative. To be more
explicit, consider a Feynman graph with $n$ external lines and $l$ loops.
Introduce a cutoff $\Lambda$ in the momentum integration to estimate the order
of divergence,%
\[
\Gamma^{\left(  n\right)  }\left(  p_{1},\cdots,p_{n-1}\right)  =\int
_{0}^{\Lambda}d^{4}q_{1}\cdots d^{4}q_{i}I\left(  p_{1},\cdots,p_{n-1}%
;q_{1},\cdots,q_{i}\right)
\]
Take a subset $S=\{q_{1}^{\prime},q_{2}^{\prime},\cdots q_{m}^{\prime}\}$ of
the loop momenta $\{q_{1},\cdots,q_{i}\}$ and scale them to infinity and all
other momenta fixed. Let $D\left(  S\right)  $ be the superficial degree of
divergence associated with integration over this set, i. e.,%
\[
\left\vert \int_{0}^{\Lambda}d^{4}q_{1}^{\prime}\cdots d^{4}q_{m}^{\prime
}I\right\vert \leq\Lambda^{D\left(  s\right)  }\left\{  \ln\Lambda\right\}
\]
where $\left\{  \ln\Lambda\right\}  $ is some function of $\ln\Lambda.$ Then
the convergent theorem states that the integral over $\{q_{1},\cdots,q_{i}\}$
converges if the $D\left(  S\right)  ^{\prime}s$ for all possible choice of
$S$ are negative. For example the graph in the following figure
\begin{figure}
\begin{center}
{\includegraphics{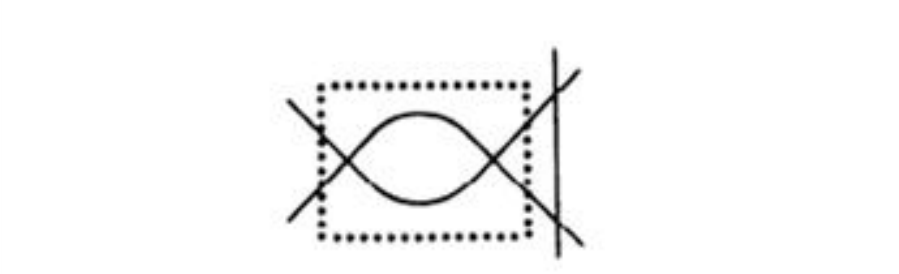}}
\\
Fig 7 divergence in 6-point function
\end{center}
\end{figure}

is a 6-point function with $D=-2.$ But the integration inside the box with
$D=0$ is logarithmically divergent. However, in the BPH procedure this
subdivergence is in fact removed by lower order counter terms as shown below.
\begin{figure}
\begin{center}
{\includegraphics{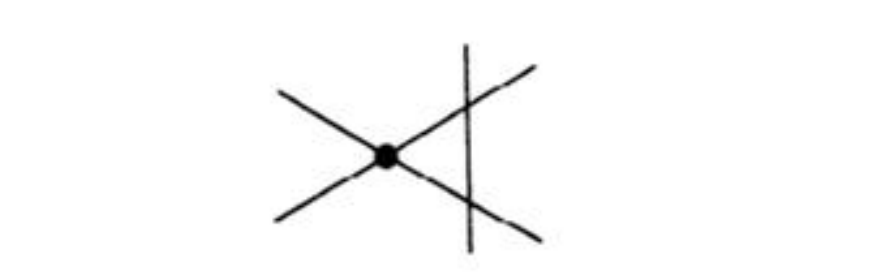}}
\\
Fig 8 Counterterm for 6-point function
\end{center}
\end{figure}

\item \underline{\textbf{Classification of divergent graphs}}\newline It is
useful to distinguish divergent graphs with different topologies in the
construction of counterterms.

\begin{enumerate}
\item \underline{Primitively divergent graphs}\newline A primitively divergent
graph has a nonnegative overall superficial degree of divergence but is
convergent for all subintegrations. Thus these are diagrams in which the only
divergences is caused by all of the loop momenta growing large together. This
means that when we differentiate with respect to external momenta at least one
of the internal loop momenta will have more power in the denominator and will
improve the convergence of the diagram. It is then clear that all the
divergences can be isolated in the first few terms of the Taylor expansion.

\item \underline{Disjointed divergent graphs}\newline Here the divergent
subgraphs are disjointed. \ For illustration, consider the 2-loop graph given
below,
\begin{figure}
\begin{center}
{\includegraphics{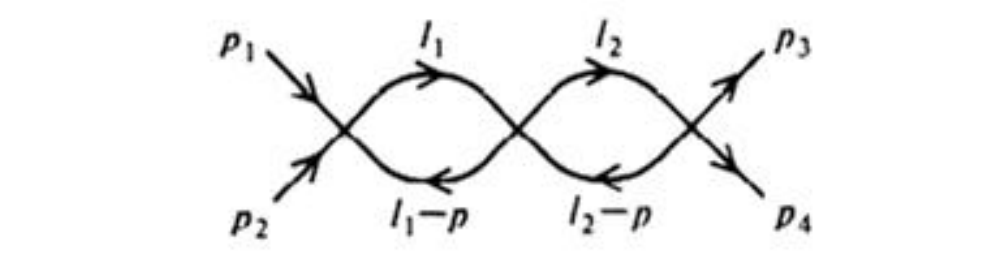}}
\\
Fig 9 two-loop disjoint divergence
\end{center}
\end{figure}

It is clear that differentiating with respect to the external momentum will
improve only one of the loop integration but not both. As a result, not all
divergences in this diagram can be removed by subtracting out the first few
terms in the Taylor expansion around external momenta. However, the lower
order counter terms in the BPH scheme will come in to save the day. The
Feynman integral is written as%
\[
\Gamma_{a}^{\left(  4\right)  }\left(  p\right)  \propto\lambda^{3}\left[
\Gamma\left(  p\right)  \right]  ^{2}%
\]
with%
\[
\Gamma\left(  p\right)  =\dfrac{1}{2}\int d^{4}l\dfrac{1}{l^{2}-\mu
^{2}+i\varepsilon}\dfrac{1}{\left[  \left(  l-p\right)  ^{2}-\mu
^{2}+i\varepsilon\right]  }%
\]
and $p=p_{1}+p_{2}.$ Since $\Gamma\left(  p\right)  $ is logarithmic
divergent, $\Gamma_{a}^{\left(  4\right)  }\left(  p\right)  $ cannot be made
convergent no matter how many derivatives act on it, even though the overall
superficial degree of divergence is zero. However, we have the lower order
counterm $-\lambda^{2}\Gamma\left(  0\right)  $ corresponding to the
substraction introduced at the 1-loop level. This generates the additional
contributions given in the following diagrams,
\begin{figure}
\begin{center}
{\includegraphics{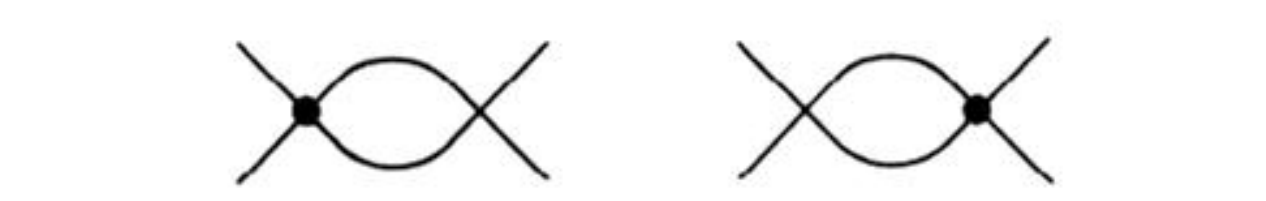}}
\\
Fig 10 two-loop graphs with counterms
\end{center}
\end{figure}

which are proportional to $-\lambda^{3}\Gamma\left(  0\right)  \Gamma\left(
p\right)  .$ Adding these 3 contributions, we get%
\begin{align*}
&  \lambda^{3}\left[  \Gamma\left(  p\right)  \right]  ^{2}-2\lambda^{3}%
\Gamma\left(  0\right)  \Gamma\left(  p\right) \\
&  =\lambda^{3}\left[  \Gamma\left(  p\right)  -\Gamma\left(  0\right)
\right]  ^{2}-\lambda^{3}\left[  \Gamma\left(  0\right)  \right]  ^{2}%
\end{align*}
Since the combination in the first $\left[  \cdots\right]  $ is finite, the
divergence in the last term can be removed by one differentiation. Here we see
that with the inclusion of lower order counterterms, the divergences take the
form of polynomials in external momenta. Thus for graphs with disjointed
divergences we need to include the lower order counter terms to remove the
divergences by substractions in Taylor expansion.

\item \underline{Nested divergent graphs}

In this case one of a pair of divergent 1PI is entirely contained within the
other as shown in the following diagram,
\begin{figure}
\begin{center}
{\includegraphics{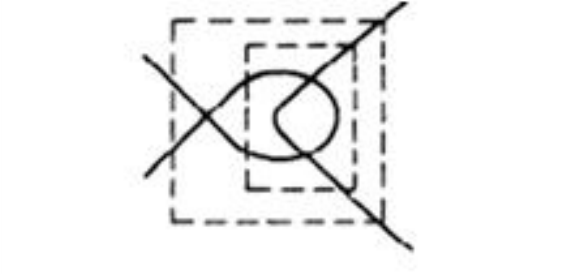}}
\\
Fig 11 Nested divergences
\end{center}
\end{figure}

After the subgraph divergence is removed by diagrams with lower order
counterterms, the overall divergences is then renormalized by a $\lambda^{3} $
counter terms as shown below,
\begin{figure}
\begin{center}
{\includegraphics{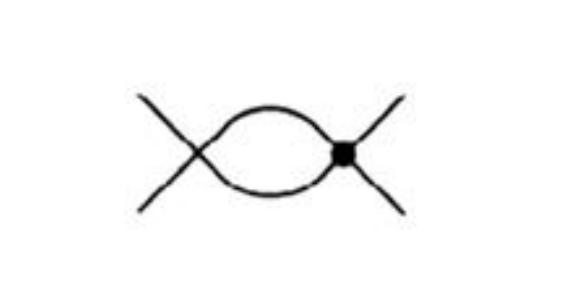}}
\\
Fig 12 lower order counterterm
\end{center}
\end{figure}

Again diagrams with lower-order counterterm insertions must be included in
order to aggregate the divergences into the form of polynomial in external momenta.

\item \underline{Overlapping divergent graphs}\newline These diagrams are
those divergences which are neither nested nor disjointed. These are most
difficult to analyze. An example of this is shown below,
\begin{figure}
\begin{center}
{\includegraphics{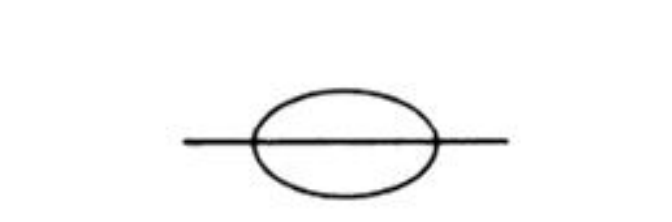}}
\\
Fig 13 Overlapping divergences
\end{center}
\end{figure}

The study of \ how to disentangle these overlapping divergences is beyond the
scope of this simple introduction and we refer interested readers to the
literature (\cite{BPH}.\cite{Zimmermann}).
\end{enumerate}
\end{enumerate}

From these discussion, it is clear that BPH renormalization scheme is quite
useful in organizing the higher order divergences in a more systematic way for
the removing of divergences by constructing the counterterms.

The general analysis of the renormalization program has been carried out by
Bogoliubov, Parasiuk, Hepp (\cite{BPH}). The result is known as BPH theorem,
which states that for a general renormalizable field theory, to any order in
perturbation theory, all divergences are removed by the counterterms
corresponding to superficially divergent amplitudes.

\section{Power counting and Renormalizability}

We now discuss the problem of renormalization for more general interactions.
It is clear that it is advantageous to use the BPH scheme in this discussion.

\subsection{Theories with fermions and scalar fields}

We first study the simple case with fermion $\psi$ and scalar field $\phi.$
Write the Lagrangian density as%
\[%
%TCIMACRO{\tciLaplace}%
%BeginExpansion
\mathcal{L}%
%EndExpansion
=%
%TCIMACRO{\tciLaplace}%
%BeginExpansion
\mathcal{L}%
%EndExpansion
_{0}+\sum_{i}%
%TCIMACRO{\tciLaplace}%
%BeginExpansion
\mathcal{L}%
%EndExpansion
_{i}%
\]
where $%
%TCIMACRO{\tciLaplace}%
%BeginExpansion
\mathcal{L}%
%EndExpansion
_{0}$ is the free Lagrangian quadratic in the fields and $%
%TCIMACRO{\tciLaplace}%
%BeginExpansion
\mathcal{L}%
%EndExpansion
_{i}$ are the interaction terms e.g.%
\[%
%TCIMACRO{\tciLaplace}%
%BeginExpansion
\mathcal{L}%
%EndExpansion
_{i}=g_{1}\overset{\_}{\psi}\gamma^{\mu}\psi\partial_{\mu}\phi,\qquad
g_{2}\left(  \overset{\_}{\psi}\psi\right)  ^{2},\qquad g_{3}\left(
\overset{\_}{\psi}\psi\right)  \phi,\qquad\cdots
\]
Here $\psi$ denotes a fermion field and $\phi$ a scalar field. Define the
following quantities%
\[%
\begin{array}
[c]{l}%
n_{i}=\text{number of }i-th\text{ type vertices}\\
b_{i}=\text{number of scalar lines in }i-th\text{ type vertex}\\
f_{i}=\text{number of fermion lines in }i-th\text{ type vertex}\\
d_{i}=\text{number of derivatives in }i-th\text{ type of vertex}\\
B=\text{number of external scalar lines}\\
F=\text{number of external fermion lines}\\
IB=\text{number of internal scalar lines}\\
IF=\text{number of internal fermion lines}%
\end{array}
\]
Counting the scalar and fermion lines, we get%
\begin{equation}
B+2\left(  IB\right)  =\sum_{i}n_{i}b_{i} \label{b-lines}%
\end{equation}%
\begin{equation}
F+2\left(  IF\right)  =\sum_{i}n_{i}f_{i} \label{f-lines}%
\end{equation}
Using momentum conservation at each vertex we can compute the number of loop
integration $L$ as%
\[
L=\left(  IB\right)  +\left(  IF\right)  -n+1,\qquad n=\sum_{i}n_{i}%
\]
where the last term is due to the overall momentum conservation which does not
contain the loop integrations. The superficial degree of divergence is then
given by%
\[
D=4L-2\left(  IB\right)  -\left(  IF\right)  +\sum_{i}n_{i}d_{i}%
\]
Using the relations given in Eqs(\ref{b-lines},\ref{f-lines}) we get%
\begin{equation}
D=4-B-\dfrac{3}{2}F+\sum_{i}n_{i}\delta_{i} \label{SuperD}%
\end{equation}
where%
\[
\delta_{i}=b_{i}+\dfrac{3}{2}f_{i}+d_{i}-4
\]
is called the $index$ $of$ $divergence$ of the interaction. Using the fact
that Lagrangian density $%
%TCIMACRO{\tciLaplace}%
%BeginExpansion
\mathcal{L}%
%EndExpansion
$ has dimension $4$ and scalar field, fermion field and the derivative have
dimensions, $1,$ $\dfrac{3}{2},$ and $1 $ respectively, we get for the
dimension of the coupling constant $g_{i}$ as%
\[
\dim\left(  g_{i}\right)  =4-b_{i}-\dfrac{3}{2}f_{i}-d_{i}=-\delta_{i}%
\]
We distinguish 3 different situations;

\begin{enumerate}
\item $\delta_{i}<0$\newline In this case, $D$ decreases with the number of
i-th type of vertices and the interaction is called $super-renormalizable$
$interaction$. The divergences occur only in some lower order diagrams. There
is only one type of theory in this category, namely $\phi^{3}$ interaction.

\item $\delta_{i}=0$\newline Here $D$ is independent of the number of i-th
type of vertices and interactions are called $renormalizable$ $interactions$.
The divergence are present in all higher-order diagrams of a finite number of
Green's functions. Interactions in this category are of the form, $g\phi^{4},$
$f\left(  \overset{\_}{\psi}\psi\right)  \phi.$

\item $\delta_{i}>0$\newline Then $D$ increases with the number of the i-th
type of vertices and all Green's functions are divergent for large enough
$n_{i}.$ These are called $non-renormalizable$ $interactions.$ There are
plenty of examples in this category, $g_{1}\overset{\_}{\psi}\gamma^{\mu}%
\psi\partial_{\mu}\phi,$ $g_{2}\left(  \overset{\_}{\psi}\psi\right)  ^{2},$
$g_{3}\phi^{5},\cdots$
\end{enumerate}

The index of divergence $\delta_{i}$ can be related to the operator's
$canonical$ $dimension$ which is defined in terms of the high energy behavior
in the free field theory. More specifically, for any operator $A,$ we write
the 2-point function as%
\[
D_{A}\left(  p^{2}\right)  =\int d^{4}xe^{-ip\cdot x}\left\langle 0\left\vert
T\left(  A\left(  x\right)  A\left(  0\right)  \right)  \right\vert
0\right\rangle
\]
If the asymptotic behavior is of the form,%
\[
D_{A}\left(  p^{2}\right)  \longrightarrow\left(  p^{2}\right)  ^{-\omega
_{A}/2},\qquad as\qquad p^{2}\longrightarrow\infty
\]
then the canonical dimension is defined as%
\[
d\left(  A\right)  =\left(  4-\omega_{A}\right)  /2
\]
Thus for the case of fermion and scalar fields we have,%
\[
d\left(  \phi\right)  =1,\qquad d\left(  \partial^{n}\phi\right)  =1+n
\]%
\[
d\left(  \psi\right)  =\dfrac{3}{2},\qquad d\left(  \partial^{n}\psi\right)
=\dfrac{3}{2}+n
\]
Note that in these simple cases, these values are the same as those obtained
in the dimensional analysis in the classical theory and sometimes they are
also called the naive dimensions. As we will see later for the vector field,
the canonical dimension is not necessarily the same as the naive dimension.

For composite operators that are polynomials in the scalar or fermion fields
it is difficult to know their asymptotic behavior. So we define their
canonical dimensions as the algebraic sum of their constituent fields. For
example,
\[
d\left(  \phi^{2}\right)  =2,\qquad d\left(  \overset{\_}{\psi}\psi\right)
=3
\]
For general composite operators that show up in the those interaction
described before, we have,%
\[
d\left(
%TCIMACRO{\tciLaplace}%
%BeginExpansion
\mathcal{L}%
%EndExpansion
_{i}\right)  =b_{i}+\dfrac{3}{2}f_{i}+d_{i}%
\]
and it is related to the index of divergence as%
\[
\delta_{i}=d\left(
%TCIMACRO{\tciLaplace}%
%BeginExpansion
\mathcal{L}%
%EndExpansion
_{i}\right)  -4
\]
We see that a dimension 4 interaction is renormalizable and greater than 4 is
non-renormalizable.\newline\underline{\textbf{Counter terms}}

Recall that we add counterterms to cancel all the divergences in Green's
functions with superficial degree of divergences $D\geq0.$ For convenience we
use the Taylor expansion around zero external momenta $p_{i}=0.$ It is easy to
see that a general diagram with $D\geq0,$ counter terms will be of the form%
\[
O_{ct}=\left(  \partial_{\mu}\right)  ^{\alpha}\left(  \psi\right)
^{F}\left(  \phi\right)  ^{B},\qquad\alpha=1,2,\cdots D
\]
and the canonical dimension is%
\[
d_{ct}=\dfrac{3}{2}F+B+\alpha
\]
The index of divergence of the counterterms is%
\[
\delta_{ct}=d_{ct}-4
\]
Using the relation in Eq (\ref{SuperD}) we can write this as%
\[
\delta_{ct}=\left(  \alpha-D\right)  +\sum_{i}n_{i}\delta_{i}%
\]
Since $\alpha\leq D$ , we have the result%
\[
\delta_{ct}\leq\sum_{i}n_{i}\delta_{i}%
\]
Thus, the counterterms induced by a Feynman diagrams have indices of
divergences less or equal to the sum of the indices of divergences of all
interactions $\delta_{i}$ in the diagram.

We then get the general result that the renormalizable interactions which have
$\delta_{i}=0$ will generate counterterms with $\delta_{ct}\leq0.$ Thus if all
the $\delta_{i}\leq0$ terms are present in the original Lagrangain, so that
the counter terms have the same structure as the interactions in the original
Lagragian, they may be considered as redefining parameters like masses and
coupling constants in the theory. On the other hand non-renormalizable
interactions which have $\delta_{i}>0$ will generate counterterms with
arbitrary large $\delta_{ct}$ in sufficiently high orders and clearly cannot
be absorbed into the original Lagrangian by a redefinition of parameters
$\delta_{ct}.$ Thus non-renormalizable theories will not necessarily be
infinite; however the infinite number of counterterms associated with a
non-renormalizable interaction will make it lack in predictive power and hence
be unattractive, in the framework of perturbation theory.

We will adopt a more restricted definition of renormalizability: a Lagrangian
is said to be renormalizable by power counting if all the counterterms induced
by the renormalization procedure can be absorbed by redefinitions of
parameters in the Lagrangian. With this definition the theory with Yukawa
interaction $\overset{\_}{\psi}\gamma_{5}\psi\phi$ by itself, is not
renormalizable even though the coupling constant is dimensionless. This is
because the 1-loop diagram shown below
\begin{figure}
\begin{center}
{\includegraphics{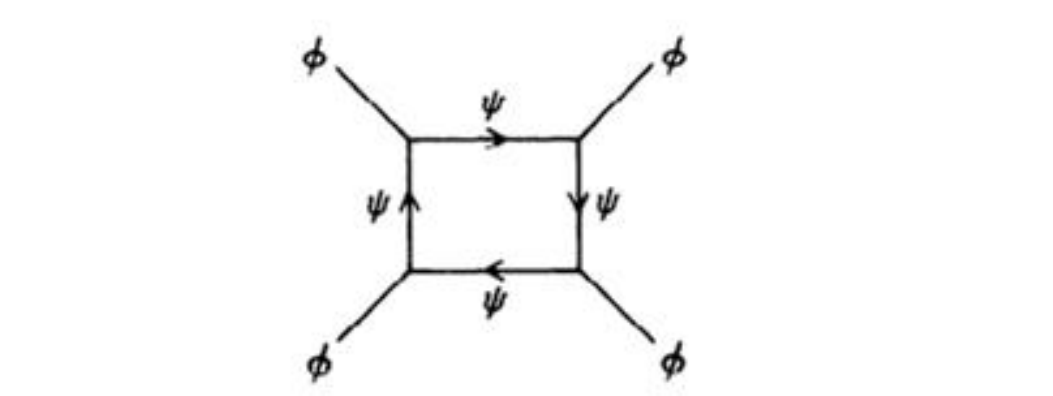}}
\\
Fig 14 Box diagram for Yukawa coupling
\end{center}
\end{figure}

is logarithmically divergent and needs a counter term of the form $\phi^{4}$
which is not present in the original Lagrangain. Thus Yukawa interaction with
additional $\phi^{4}$ interaction is renormalizable.\newline\underline
{\textbf{Theories with vector fields}}

Here we distinguish massless from massive vector fields because their
asymptotic behaviors for the free field propagators are very different.

\begin{enumerate}
\item \underline{Massless vector field}\newline Massless vector field is
usually associated with local gauge invariance as in the case of QED. The
asymptotic behavior of free field propagator for such vector field is very
similar to that of scalar field. For example, in the Feynman gauge we have%
\[
\Delta_{\mu\nu}\left(  k\right)  =\dfrac{-ig_{\mu\nu}}{k^{2}+i\varepsilon
}\longrightarrow O\left(  k^{-2}\right)  ,\qquad\text{for large }k^{2}%
\]
which has the same asymptotic behaior as that of scalar field. Thus the power
counting for theories with massless vector field interacting with fermions and
scalar fields is the same as before. The renomalizable interactions in this
category are of the type,%
\[
\overset{\_}{\psi}\gamma_{\mu}\psi A^{\mu},\qquad\phi^{2}A_{\mu}A^{\mu}%
,\qquad\left(  \partial_{\mu}\phi\right)  \phi A^{\mu}%
\]
Here $A^{\mu}$ is a massless vector field and $\psi$ a fermion field.

\item \underline{Massive vector field}\newline Here the free Lagrangian is of
the form,%
\[%
%TCIMACRO{\tciLaplace}%
%BeginExpansion
\mathcal{L}%
%EndExpansion
_{0}=-\dfrac{1}{4}\left(  \partial_{\mu}V_{\nu}-\partial_{\nu}V_{\mu}\right)
^{2}+\dfrac{1}{2}M_{V}^{2}V_{\mu}^{2}%
\]
where $V_{\mu}$ is a massive vector field and $M_{V}$ is the mass of the
vector field. The propagator in momentum space is of the form,%
\begin{equation}
D_{\mu\nu}\left(  k\right)  =\dfrac{-i\left(  g_{\mu\nu}-k_{\mu}k_{\nu}%
/M_{V}^{2}\right)  }{k^{2}-M_{V}^{2}+i\varepsilon}\longrightarrow O\left(
1\right)  ,\text{ \ \ as \ }k\rightarrow\infty\label{Massive V}%
\end{equation}
This implies that canonical dimension of massive vector field is two rather
than one. The power counting is now modified with superficial degree of
divergence given by%
\[
D=4-B-\dfrac{3}{2}F-V+\sum_{i}n_{i}\left(  \Delta_{i}-4\right)
\]
with%
\[
\Delta_{i}=b_{i}+\dfrac{3}{2}f_{i}+2v_{i}+d_{i}%
\]
Here $V$ is the number of external vector lines, $v_{i}$ is the number of
vector fields in the $i$th type of vertex and $\Delta_{i}$ is the canonical
dimension of the interaction term in $%
%TCIMACRO{\tciLaplace}%
%BeginExpansion
\mathcal{L}%
%EndExpansion
$. From the formula for $\Delta_{i}$ we see that the only renormalizable
interaction involving massive vector field, $\Delta_{i}\leq4,$ is of the form,
$\phi^{2}A_{\mu} $ and is not Lorentz invariant. Thus there is no nontrivial
interaction of the massive vector field which is renormalizabel. However, two
important exceptions should be noted;

\begin{enumerate}
\item In a gauge theory with spontaneous symmetry breaking, the gauge boson
will acquire mass in such a way to preserve the renormalizability of the
theory (\cite{Hooft}).

\item A theory with a neutral massive vector boson coupled to a conserved
current is also renormalizable. Heuristically, we can understand this as
follows. The propagator in Eq(\ref{Massive V}) always appears between
conserved currents $J^{\mu}\left(  k\right)  $ and $J^{\nu}\left(  k\right)  $
and the $k_{\mu}k_{\nu}/M_{V}^{2}$ term will not contribute because of current
conservation, $k^{\mu}J_{\mu}\left(  k\right)  =0$ or in the coordinate space
$\partial^{\mu}J_{\mu}\left(  x\right)  =0.$ Then the power counting is
essentially the same as for the massless vector field case.
\end{enumerate}
\end{enumerate}

\bigskip

\subsection{Renormalization of Composite Operators}

In some cases, we need to consider the Green's function of composite operator,
an operator with more than one fields at the same space time. Consider a
simple composite operator of the form $\Omega(x)=\frac{1}{2}\phi^{2}(x)$ in
$\lambda\phi^{4}$ theory. Green's function with one insertion of $\Omega$ is
of the form,
\[
G_{\Omega}^{(n)}(x;x_{1},x_{2},x_{3},...,x_{n})=\left\langle 0|T(\frac{1}%
{2}\phi^{2}(x)\phi(x_{1})\phi(x_{2})...\phi(x_{n}))|0\right\rangle
\]
In momentum space we have
\begin{align*}
&  (2\pi)^{4}\delta^{4}(p+p_{1}+p_{2}+...+p_{n})G_{\phi^{2}}^{(n)}%
(p;p_{1},p_{2},p_{3},...,p_{n})\\
&  =\int d^{4}x\;e^{-ipx}\int%
%TCIMACRO{\dprod \limits_{i=1}^{n}}%
%BeginExpansion
{\displaystyle\prod\limits_{i=1}^{n}}
%EndExpansion
d^{4}x_{i}e^{-ip_{i}x_{i}}G_{\Omega}^{(n)}(x;x_{1},x_{2},x_{3},...,x_{n})
\end{align*}
In perturbation theory, we can use Wick's theorem (\cite{Wick}) to work out
these Green's functions in terms of Feynman diagram.\newline Example, to
lowest order in $\lambda$ the 2-point function with one composite operator
$\Omega(x)=\frac{1}{2}\phi^{2}(x)$ is, after using the Wick's theorem,
\[
G_{\phi^{2}}^{(2)}(x;x_{1},x_{2})=\frac{1}{2}\left\langle 0|T\{\phi^{2}%
(x)\phi(x_{1})\phi(x_{2})\}|0\right\rangle =i\Delta(x-x_{1})i\Delta(x-x_{2})
\]
or in momentum space
\[
G_{\phi^{2}}^{(2)}(p;p_{1},p_{2})=i\Delta(p_{1})i\Delta(p+p_{1})
\]
If we truncate the external propagators, we get
\[
\Gamma_{\phi^{2}}^{(2)}(p,p_{1},-p_{1}-p)=1
\]
\begin{figure}
\begin{center}
{\includegraphics{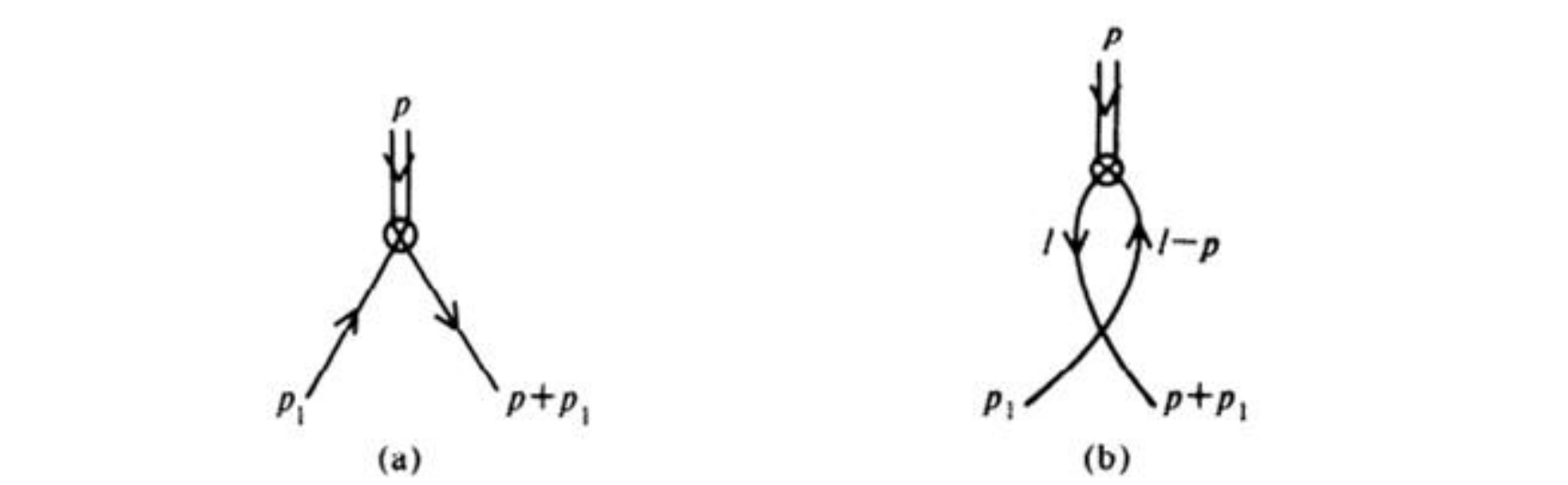}}
\\
Fig 15 Graphs for composite operator
\end{center}
\end{figure}

To first order in $\lambda$, we have
\[
G_{\phi^{2}}^{(2)}(x,x_{1},x_{2})=\int\left\langle 0|T\{\frac{1}{2}\phi
^{2}(x)\phi(x_{1})\phi(x_{2})\frac{\left(  -i\lambda\right)  }{4!}\phi
^{4}(y)\}|0\right\rangle d^{4}y
\]%
\[
=\int d^{4}y\frac{-i\lambda}{2}[i\Delta(x-y)]^{2}i\Delta(x_{1}-y)i\Delta
(x_{2}-y)
\]
The amputated 1PI momentum space Green's function is
\[
\Gamma_{\phi^{2}}^{(2)}(p;p_{1},-p-p_{1})=\frac{-i\lambda}{2}\int\frac{d^{4}%
l}{(2\pi)^{4}}\frac{i}{l^{2}-\mu^{2}+i\epsilon}\frac{i}{(l-p)^{2}-\mu
^{2}+i\epsilon}%
\]
To calculate this type of Green's functions systematically, we can add a term
$\chi(x)\Omega(x)$ to
%TCIMACRO{\tciLaplace}%
%BeginExpansion
$\mathcal{L}$%
%EndExpansion
\
\[%
%TCIMACRO{\tciLaplace}%
%BeginExpansion
\mathcal{L}%
%EndExpansion
\lbrack\chi]=%
%TCIMACRO{\tciLaplace}%
%BeginExpansion
\mathcal{L}%
%EndExpansion
\lbrack0]+\chi(x)\Omega(x)
\]
where $\chi(x)$ is a c-number source function. We can construct the generating
functional $W[\chi]$ in the presence of this external source. We obtain the
connected Green's function by differentiating $\ln W[\chi]$ with respect to
$\chi$ and then setting $\chi$ to zero.\newline\underline
{\textbf{Renormalization of composite operators}}\newline Superficial degrees
of divergence for Green 's function with one composite operator is,%
\[
D_{\Omega}=D+\delta_{\Omega}=D+(d_{\Omega}-4)
\]
where $d_{\Omega}$ is the canonical dimension of $\Omega$. For the case of
$\Omega(x)=\frac{1}{2}\phi^{2}(x),\quad d_{\phi^{2}}=2\;$and$\;D_{\phi^{2}%
}=2-n\Rightarrow$ only $\Gamma_{\phi^{2}}^{(2)}$ is divergent. Taylor
expansion takes the form,\newline%
\[
\Gamma_{\phi^{2}}^{(2)}(p;p_{1})=\Gamma_{\phi^{2}}^{(2)}(0,0)+\Gamma_{\phi
^{2}R}^{(2)}(p,p_{1})
\]
We can combine the counter term
\[
\frac{-i}{2}\Gamma^{(2)}{\phi^{2}}(0,0)\chi(x)\phi^{2}(x)
\]
with the original term to write
\[
\frac{-i}{2}\chi\phi-\frac{i}{2}\Gamma_{\phi^{2}}^{2}(0,0)\chi\phi^{2}%
=-\frac{i}{2}Z_{\phi^{2}}\chi\phi^{2}%
\]
In general, we need to insert a counterterm $\Delta\Omega$ into the original
addition
\[
L\rightarrow L+\chi(\Omega+\Delta\Omega)
\]
If $\Delta\Omega=C\Omega$, as in the case of $\Omega=\frac{1}{2}\phi^{2} $, we
have
\[
L[\chi]=L[0]+\chi Z_{\Omega}\Omega=L[0]+\chi\Omega_{0}%
\]
with
\[
\Omega_{0}=Z_{\Omega}\Omega=(1+C)\Omega
\]
Such composite operators are said to be multiplicative renormalizable and
Green's functions of unrenormalized operator $\Omega_{0}$ is related to that
of renormalized operator $\Omega$ by
\[
G_{\Omega_{0}}^{(n)}(x;x_{1},x_{2},...x_{n})=\left\langle 0|T\{\Omega
_{0}(x)\phi(x_{1})\phi(x_{2})...\phi(x_{n})\}|0\right\rangle
\]%
\begin{equation}
\quad\quad\quad=Z_{\Omega}Z_{\phi}^{n/2}G_{lR}^{(n)}(x;x_{1},...x_{n})
\label{Composite}%
\end{equation}
For more general cases, $\Delta\Omega\neq c\Omega$ and the renormalization of
a composite operator may require counterterms proportional to other composite
operators.\newline Example: Consider 2 composite operators $A$ and $B.$ Denote
the counterterms by $\Delta A$ and $\Delta B.$ Including the counter terms we
can write,\newline%
\[
L[\chi]=L[0]+\chi_{A}(A+\Delta A)+\chi_{B}(B+\Delta B)
\]
Very often with counterterms $\Delta A$ and $\Delta B$ are linear combinations
of A and B\newline%
\[
\Delta A=C_{AA}A+C_{AB}B
\]%
\[
\Delta B=C_{BA}A+C_{BB}B
\]
We can write
\[
L[\chi]=L[0]+\left(  \chi_{A}\;\chi_{B}\right)  \{C\}\left(
\begin{array}
[c]{c}%
A\\
B
\end{array}
\right)  \quad\text{where}\;\;\{C\}=\left(
\begin{array}
[c]{cc}%
1+C_{AA} & C_{AB}\\
C_{BA} & 1+C_{BB}%
\end{array}
\right)
\]
Diagonalize $\{C\}$ by bi-unitary transformation
\[
U\{C\}V^{+}=\left(
\begin{array}
[c]{cc}%
Z_{A^{^{\prime}}} & 0\\
0 & Z_{B^{^{\prime}}}%
\end{array}
\right)
\]
Then
\[
L[\chi]=L[0]+Z_{A^{^{\prime}}}\chi_{A^{^{\prime}}}A^{^{\prime}}+Z_{B^{^{\prime
}}}\chi_{B^{^{\prime}}}B^{^{\prime}}%
\]%
\[
\left(
\begin{array}
[c]{c}%
A^{^{\prime}}\\
B^{^{\prime}}%
\end{array}
\right)  =V\left(
\begin{array}
[c]{c}%
A\\
B
\end{array}
\right)  \quad\left(  \chi_{A^{^{\prime}}}\ \chi_{B^{^{\prime}}}\right)
=\left(  \chi_{A}\;\ \chi_{B}\right)  U
\]
and $A^{^{\prime}}\;,B^{^{\prime}}$ are multiplicatively
renormalizable.\newline

\subsection{Symmetry and Renormalization}

For a theory with global symmetry, we require that the counter terms should
also respect the symmetry. For example, consider the Lagrangian given by%
\begin{equation}%
%TCIMACRO{\tciLaplace}%
%BeginExpansion
\mathcal{L}%
%EndExpansion
=\dfrac{1}{2}\left[  \left(  \partial_{\mu}\phi_{1}\right)  ^{2}+\left(
\partial_{\mu}\phi_{2}\right)  ^{2}\right]  -\dfrac{\mu^{2}}{2}\left(
\phi_{1}^{2}+\phi_{2}^{2}\right)  -\dfrac{\lambda}{4}\left(  \phi_{1}^{2}%
+\phi_{2}^{2}\right)  ^{2} \label{O(2) symmetry}%
\end{equation}
This Lagrangian has the $O\left(  2\right)  $ symmetry given below
\[
\phi_{1}\rightarrow\phi_{1}^{\prime}=\cos\theta\phi_{1}+\sin\theta\phi
_{2}\text{ \ \ }%
\]%
\[
\phi_{2}\rightarrow\phi_{2}^{\prime}=-\sin\theta\phi_{1}+\cos\theta\phi_{2}%
\]
The counter terms for this theory should have the same symmetry. For example
the mass counter term should be of the form%
\[
\delta\mu^{2}\left(  \phi_{1}^{2}+\phi_{2}^{2}\right)
\]
i.e. the coefficient of $\phi_{1}^{2}$ counter term should be the same as
$\phi_{2}^{2}$ term. Then the only other possible counter terms are of the
form,%
\[
\left(  \partial_{\mu}\phi_{1}\right)  ^{2}+\left(  \partial_{\mu}\phi
_{2}\right)  ^{2},\qquad\left(  \phi_{1}^{2}+\phi_{2}^{2}\right)  ^{2}%
\]

\begin{enumerate}
\item \underline{\textbf{Broken symmetry and renormalization}}\newline For the
case the symmetry is slightly broken an interesting feature occurs. We will
illustrate this with a simple case where the symmetry breaking is of the form,%
\[%
%TCIMACRO{\tciLaplace}%
%BeginExpansion
\mathcal{L}%
%EndExpansion
_{SB}=c\left(  \phi_{1}^{2}-\phi_{2}^{2}\right)
\]
Since the index of divergence for $%
%TCIMACRO{\tciLaplace}%
%BeginExpansion
\mathcal{L}%
%EndExpansion
_{SB}$ is $\delta_{SB}=-2,$ the superficial degree of divergence for graphs
containing $%
%TCIMACRO{\tciLaplace}%
%BeginExpansion
\mathcal{L}%
%EndExpansion
_{SB} $ is
\[
D_{SB}=4-B_{1}-B_{2}-2n_{SB}%
\]
where $B_{1},B_{2}$ are number of external $\phi_{1},\phi_{2}$ lines and
$n_{SB}$ is the number of times $%
%TCIMACRO{\tciLaplace}%
%BeginExpansion
\mathcal{L}%
%EndExpansion
_{SB}$ appears in the graph. For the case $n_{SB}=1,$ we have%
\[
D_{SB}=2-B_{1}-B_{2}%
\]
This means that $D_{SB}\geq0$ only for $B_{1}=2$, $B_{2}=0,$ or $B_{1}=0$,
$B_{2}=2$ and the counter terms we need are $\phi_{1}^{2}$, and $\phi_{2}%
^{2}.$ The combination $\phi_{1}^{2}+\phi_{2}^{2}$ can be absorbed in the mass
counter term while the other combination $\phi_{1}^{2}-\phi_{2}^{2}$ can be
absorbed into $%
%TCIMACRO{\tciLaplace}%
%BeginExpansion
\mathcal{L}%
%EndExpansion
_{SB}.$ This shows the when the symmetry is broken, the counterterms we need
will have the property that,%
\[
\delta_{CT}\leq\delta_{SB}%
\]
Or in terms of operator dimension%
\[
\dim\left(
%TCIMACRO{\tciLaplace}%
%BeginExpansion
\mathcal{L}%
%EndExpansion
_{CT}\right)  \leq\dim\left(
%TCIMACRO{\tciLaplace}%
%BeginExpansion
\mathcal{L}%
%EndExpansion
_{SB}\right)
\]
Thus when $\dim\left(
%TCIMACRO{\tciLaplace}%
%BeginExpansion
\mathcal{L}%
%EndExpansion
_{SB}\right)  \leq3,$ the dimension of counter terms cannot be $4.$ This
situation is usually referred to as soft breaking of the symmetry. This is
known as the $Szymanzik$ $theorem$ (\cite{Szymanzik}). Note that for the soft
breaking the coupling constant $g_{SB}$ will have positive dimension of mass
and will be negligible when energies become much larger than $g_{SB}.$ In
other words, the symmetry will be restored at high energies.

\item \underline{\textbf{Ward Identity}} (\cite{Ward})\newline In case of
global symmetry, we also have some useful relation for composite operator like
the current operator which generates the symmetry. We will give a simple
illustration of this feature. The Lagrangian given in Eq (\ref{O(2) symmetry})
can be rewritten as%
\[%
%TCIMACRO{\tciLaplace}%
%BeginExpansion
\mathcal{L}%
%EndExpansion
=\partial_{\mu}\phi^{\dag}\partial_{\mu}\phi-\mu^{2}\phi^{\dag}\phi
-\lambda\left(  \phi^{\dag}\phi\right)  ^{2}%
\]
where
\[
\phi=\dfrac{1}{\sqrt{2}}\left(  \phi_{1}+i\phi_{2}\right)
\]
The symmetry transformation is then%
\[
\phi\rightarrow\phi^{\prime}=e^{i\theta}\phi
\]
This will give rise, through Noether's theorem, the current of the form,%
\[
J_{\mu}=i\left[  \left(  \partial_{\mu}\phi^{\dag}\right)  \phi-\left(
\partial_{\mu}\phi\right)  \phi^{\dag}\right]
\]
is conserved,%
\[
\partial^{\mu}J_{\mu}=0
\]
From the canonical commutation relation,%
\[
\left[  \partial_{0}\phi^{\dag}(\overset{\rightarrow}{x},t),\;\phi
(\overset{\rightarrow}{x}^{\prime},t)\right]  =-i\delta^{3}(\overset
{\rightarrow}{x}-\overset{\rightarrow}{x}^{\prime})
\]
we can derive,%
\begin{equation}
\left[  J_{0}(\overset{\rightarrow}{x},t),\;\phi(\overset{\rightarrow}%
{x}^{\prime},t)\right]  =\delta^{3}(\overset{\rightarrow}{x}-\overset
{\rightarrow}{x}^{\prime})\phi(\overset{\rightarrow}{x}^{\prime},t)
\label{comm1}%
\end{equation}%
\begin{equation}
\left[  J_{0}(\overset{\rightarrow}{x},t),\;\phi^{\dag}(\overset{\rightarrow
}{x}^{\prime},t)\right]  =-\delta^{3}(\overset{\rightarrow}{x}-\overset
{\rightarrow}{x}^{\prime})\phi^{\dag}(\overset{\rightarrow}{x}^{\prime},t)
\label{comm2}%
\end{equation}
Now consider the Green's function of the form,%
\[
G_{\mu}\left(  p,q\right)  =\int d^{4}xd^{4}ye^{-iq\cdot x-ip\cdot
y}\left\langle 0\left\vert T\left(  J_{\mu}\left(  x\right)  \phi\left(
y\right)  \phi^{\dag}\left(  0\right)  \right)  \right\vert 0\right\rangle
\]
Multiply $q^{\mu}$ into this Green's function,%
\begin{align*}
q^{\mu}G_{\mu}\left(  p,q\right)   &  =-i\int d^{4}xd^{4}ye^{-iq\cdot
x-ip\cdot y}\partial_{x}^{\mu}\left\langle 0\left\vert T\left(  J_{\mu}\left(
x\right)  \phi\left(  y\right)  \phi^{\dag}\left(  0\right)  \right)
\right\vert 0\right\rangle \\
&  =-i\int d^{4}xe^{-i(q+p)\cdot x}\left\langle 0\left\vert T\left(
\phi\left(  x\right)  \phi^{\dag}\left(  0\right)  \right)  \right\vert
0\right\rangle \\
&  +i\int d^{4}xe^{-ip\cdot y}\left\langle 0\left\vert T\left(  \phi\left(
y\right)  \phi^{\dag}\left(  0\right)  \right)  \right\vert 0\right\rangle
\end{align*}
where we have used the current conservation and commutators in Eqs
(\ref{comm1},\ref{comm2}). The right-hand side here is just the propagator for
the scalar field,%
\[
\Delta\left(  p\right)  =\int d^{4}xe^{-ip\cdot x}\left\langle 0\left\vert
T\left(  \phi\left(  x\right)  \phi^{\dag}\left(  0\right)  \right)
\right\vert 0\right\rangle
\]
and we get%
\begin{equation}
-iq^{\mu}G_{\mu}\left(  p,q\right)  =\Delta\left(  p+q\right)  -\Delta\left(
p\right)  \label{Ward}%
\end{equation}
This is example of Ward identity (\cite{Ward}).\newline This relation is
derived in terms of unrenomalized fields which satisfy the canonical
commutation relation. In terms of renormalized quantities,
\[
G_{\mu}^{R}\left(  p,q\right)  =Z_{\phi}^{-1}Z_{J}^{-1}G_{\mu}\left(
p,q\right)  ,\qquad\Delta^{R}\left(  p\right)  =Z_{\phi}^{-1}\Delta\left(
p\right)
\]
the Ward identity in Eq (\ref{Ward}) becomes%
\[
-iZ_{J}q^{\mu}G_{\mu}^{R}\left(  p,q\right)  =\Delta^{R}\left(  p+q\right)
-\Delta^{R}\left(  p\right)
\]
Since the right-hand side is cutoff independent, $Z_{J}$ on the left-hand side
is also cutoff independent, and we do not need any counter terms to
renormalize $J_{\mu}\left(  x\right)  .$ In other words, the conserved current
$J_{\mu}\left(  x\right)  $ is not renormalized as composite operator, i.e.
$Z_{J}=1$. Thus the relation for the renormalized quantities takes the simple
form,%
\[
-iq^{\mu}G_{\mu}^{R}\left(  p,q\right)  =\Delta^{R}\left(  p+q\right)
-\Delta^{R}\left(  p\right)
\]
Such a non-renormalization result holds for many conserved quantities.
\end{enumerate}

I would like to thank Professor Yungui Gong for hospitality during my visit to
Chongqing Univeristy of Posts and Telecommunications where parts of this
manuscript is written.


\bigskip

\begin{thebibliography}{99}                                                                                               

\bibitem {Renormalization}R. P. Feynman, Phys. Rev. \textbf{74}, 939, 1430
(1948), J. Schwinger , Phys. Rev. \textbf{73}, 416 (1948), \textbf{75}, 898
(1949), S. Tomonaga, Phys. Rev. \textbf{74}, 224 (1948), F. J. Dyson, Phys.
Rev. \textbf{75}, 486 (1949).

\bibitem {Renormalization-advanced}C. Itzykson, and J.-B. Zuber, "Quantum
Field Theory", McGraw-Hill, New York, (1980). N. N. Bogoliubov and D. V.
Shirkov, "Introduction to Theories of Quantized Fields" (3rd edition)
Wiley-Interscience, New York, (1980). M. E. Peskin and D. Schroeder, "An
Introduction to Quantum Field Theory", Addison-Wiley, New York, 1995. S.
Weinberg, " The Quantum Theory of Fields" Vol, 1, 2, and 3, Cambridge
University Press, Cambridge (1995).

\bibitem {BPH}N. N. Bogoliubov and O. S. Parasiuk, Acta. Math. \textbf{97},
227 (1957), K. Hepp, Comm. Math. Phys. 2, 301, (1966), W. Zimmermann, In
"Lectures on Elementary Particles and Quantum Field Theory", Proc. 1970
Brandeis Summer Institute (ed. S. Deser et al) MIT\ Press, Cambridge, Mass. (1970).

\bibitem {Zimmermann}W. Zimmermann, in "Lectures on elementary particle and
quantum field theory" Pro. 1970 Brandies Summer Institute (ed S. Deser) MIT
Press Cambridge, Massachusetts.

\bibitem {Dim reg}G. t' Hooft and M. Veltman, Nucl. Phys. \textbf{B44}, 189,
(1972), C. G. Bollini and J. J. Giambiagi, Phys. Lett. \textbf{40B, }566,
(1972), J. F. Ashmore, Nuovo Cimento Lett. \textbf{4, }289, (1972), G. M.
Cicuta and Mortaldi, Nuovo Cimento Lett. \textbf{4, }329, (1972).

\bibitem {Renor Group}C.\ G. Callan, Phys. Rev. \textbf{D2}, 1541, (1970), K.
Szymanzik, Comm. Math. Phys. \textbf{18}, 227, (1970).

\bibitem {Weinberg theorem}S. Weinberg, Phys. Rev. \textbf{118}, 838, (1960).

\bibitem {Hooft}G. 't Hooft, Nucl. Phys. \textbf{B35}, 173, (1971).

\bibitem {Wick}G. C. Wick, Phys. Rev. \textbf{80, }268 (1950)\textbf{.}

\bibitem {Szymanzik}K. Szymanzik, in Coral Gables Conf. on Fundamental
interactions at high energies II, ed. A. Perlmutter, G. J. Iverson and R.M.
Williams (Gordon and Breach, New York, (1970).

\bibitem {Ward}J. C. Ward, Phys. Rev. \textbf{78}, 1824, (1950).
\end{thebibliography}
\end{document}